\documentclass[aps,twocolumn,superscriptaddress,showpacs]{revtex4-1}
\usepackage{longtable}
\usepackage{verbatim}
\usepackage{multirow}
\usepackage{bm,graphicx}
\begin{document}
\bibliographystyle{apsrev}


\title{Multichannel parametrization of $\overline K N$ scattering amplitudes\\
        and extraction of resonance parameters}


\author{H. Zhang}
\author{J. Tulpan}
\author{M. Shrestha}
\author{D.~M.~Manley}
\affiliation{Department of Physics, Kent State University, Kent, OH 44242-0001}



\date{\today}

\begin{abstract}
We present results of a new multichannel partial-wave analysis for 
$\overline K N$ scattering in the c.m.\ energy range 1480 to 2100 MeV.
Resonance parameters were extracted by fitting partial-wave amplitudes
from all considered channels using a multichannel parametrization that
is consistent with $S$-matrix unitarity. 
The resonance parameters are generally in good agreement with predictions of the Koniuk-Isgur quark model.
\end{abstract}


\maketitle

\section{INTRODUCTION}


The study of hyperon resonances ($\Lambda^*$ and $\Sigma^*$) is motivated, in part, by our relatively poor knowledge of their properties \cite{pdg12} compared to those of non-strange baryons ($N$ and $\Delta$ resonances) and also by the quest to enhance the understanding of the role of confinement in the non-perturbative region.

Most prior partial-wave analyses \cite{Armenteros1969, Conforto1971, Horn1975_1, Hemingway1975, Baillon1975, Gopal1977} assumed a simple energy-dependent parametrization for the partial-wave amplitudes (typically, a sum of Breit-Wigner resonances plus polynomial backgrounds). Such a parametrization introduces a model-dependent bias and results in a violation of unitarity of the partial-wave $S$-matrix.
One of the disturbing features that appears when examining the partial waves obtained from these is that they do not always join smoothly with the partial waves given in analyses done for the same channel over a different energy range.
Among these works, one of the arguably more sophisticated partial-wave analyses was that done by Gopal {\it et al.} {\cite{Gopal1977}}.
They analyzed world data available at that time for $\overline{K}N \rightarrow \overline{K} N$, $\overline{K}N \rightarrow \pi \Lambda$, and $\overline{K}N \rightarrow \pi\Sigma$ in the c.m.\ energy range from 1480 to 2170 MeV.
In their analysis, parallel single-channel analyses were performed using the energy-dependent technique discussed above. That is, an attempt was made to obtain consistent resonance parameters in all reactions, but independent backgrounds were allowed in each channel.
In their work, a conventional energy-dependent analysis was performed first for each of the three two-body channels.
The presence of a resonance in a partial wave was, as usual, detected by comparing the goodness of fit.
The wave was parametrized as a smooth background to the alternative fit when a Breit-Wigner was added to the background.
The three separate fits were then considered together in order to obtain a real multichannel analysis that required the masses and widths of the resonances be the same in each channel.
The final fits were done with resonance parameters fixed and equal to a ``weighted average'' of the three values.
There are several shortcomings in the parametrization used by Gopal {\it et al.}.
Firstly, it is inconsistent with unitarity.
Secondly, it does not treat background consistently in different channels.
Thirdly, it is unable to allow for cusp effects associated with such channels as $\eta \Lambda$ and $\eta \Sigma$.
Fourthly, it fails to handle threshold effects consistently for all partial waves (through the barrier-penetration factors).
Finally, pole positions (i.e., pole mass and width) of resonances were not determined.

The main goal of the present work was to extract resonance parameters from a unitary, multichannel partial-wave analysis of $\overline{K} N$ scattering reactions.
 Our approach is different and unique in the sense that it uses a generalized energy-dependent Breit-Wigner parametrization of amplitudes treating all the channels on an equal footing and taking full account of non-resonant backgrounds. 
 The main advantage of using a multichannel partial-wave analysis to extract resonance parameters is that several reactions can be described simultaneously using a consistent set of resonance parameters for all reactions, all without violating unitarity of the partial-wave $S$-matrix.
Furthermore, small signals for a resonance in one reaction channel might be overlooked, whereas if similar signals appear in several channels, there is a good chance that the resonance can be identified and its parameters measured.

The channels included in this analysis are $\overline K N$, $\pi \Lambda$, $\pi \Sigma$, $\pi \Lambda(1520)$, $\pi \Sigma(1385)$, $\overline K^*N$, and $\overline K\Delta$. We begin with an energy-dependent model for fitting of the $\overline K N$ partial-wave data. Our detailed partial-wave analyses of reactions $\overline K N\rightarrow \overline K N$, $\overline K N\rightarrow \pi\Lambda $, and $\overline K N\rightarrow \pi\Sigma$ are presented elsewhere \cite{manoj13, hongyu08}. The reliability of the energy-dependent amplitudes extracted from this work was tested by using the fitted amplitudes to compare with various observables. Our solution was in good agreement with available data for $\overline K N\rightarrow \overline K N$, $\overline K N\rightarrow \pi\Lambda $, and $\overline K N\rightarrow \pi\Sigma$ \cite{manoj13, hongyu08}. 

\section{Theoretical Aspects}
The KSU model, developed by Manley \cite{manley03},  employs a unitary multichannel parameterization to extract resonance parameters. It has been successfully used to extract $N^*$ and $\Delta^*$ parameters from a combined fit of $\pi N\rightarrow \pi N$, $\pi N\rightarrow \pi\pi N$, $\pi N\rightarrow \eta N$, $\pi N\rightarrow K\Lambda$, and $\gamma N\rightarrow \pi N$ amplitudes \cite{manoj12}. The present work has utilized the model to extract the $\Lambda^*$ and $\Sigma^*$ parameters from multichannel fits of $\overline KN$ scattering amplitudes \cite{john07, hongyu08}.

 In the KSU model,  the partial-wave $S$-matrix is parametrized as
\begin{equation}
   S= B^TRB=I+2{\rm i}T~,
 \end{equation}
where $T$ is the corresponding partial-wave $T$-matrix. Here $R$ is a unitary, symmetric, and generalized multichannel Breit-Wigner matrix while $B$ and its transpose $B^T$ are unitary matrices describing non-resonant background. The background matrix $B$ is constructed from a product of unitary matrices: $B = B_1B_2\cdots B_n$, where $n$ is a very small interger. Further details about the amplitude parametrization can be found in Refs. \cite{hongyu08, manley03, manoj12}.

\section{Fitting Procedure}
The amplitudes for the multichannel energy-dependent fit were obtained from various partial-wave analyses.  Our energy-dependent fit included the Cameron78 solution for $\overline K N\rightarrow\pi \Sigma(1385)$ \cite{Cameron1978}, the Cameron77 solution for $\overline K N\rightarrow\pi \Lambda(1520)$ \cite{Cameron1977}, the Cameron78 solution for $\overline K N\rightarrow \overline K^*N$ \cite{Cameron1978_1}, and the Litchfield74 solution for $\overline K N\rightarrow \overline K \Delta$ \cite{Litch1974}. In addition, we included our single-energy amplitudes for $\overline K N\rightarrow \overline K N$, $\overline K N\rightarrow \pi \Lambda$, and $\overline K N\rightarrow \pi\Sigma$ \cite{manoj13}. Previous single-channel analyses \cite{Armenteros1969, Baillon1975, Horn1975_1} of $\overline K N\rightarrow \overline KN$,  $\overline K N\rightarrow \pi\Sigma$, and $\overline K N \rightarrow \pi\Lambda$ were simplistic energy-dependent PWAs that failed to satisfy $S$-matrix unitarity. A multichannel energy-dependent fit was performed in the c.m.\ energy range $W$ from 1480 to 2170 MeV. Initially some approximately known fitting parameters were held fixed to yield a good fit. In some partial waves,  $\sigma\Lambda$ and $\sigma\Sigma$ channels were included as dummy channels (channels without data) to satisfy unitarity. The $\eta\Lambda$ channel was included for the $S_{01}$ partial wave by fitting data \cite{starostin2001} for $\sigma(K^-p \rightarrow \eta\Lambda)$ up to a c.m.\ energy of 1685 MeV. In addition, the $\eta\Sigma$ channel was included as a dummy channel for the $S_{11}$ partial wave. The fitting parameters for each resonance were Breit-Wigner masses $M$, and decay amplitudes $\pm\sqrt{\Gamma_i}$, given by the signed square roots of the partial widths. In our final fits, uncertainties in resonance parameters were calculated with all fitting parameters free to vary. Uncertainties were propagated into all resonance parameters using the full error matrix for each fit.

 \section{Discussion of Resonance Parameters}
The resonance parameters for states with $I=0$ and $I=1$ are listed in Tables I and II, respectively. The first column lists the resonance name together with the resonance mass and its total width in MeV. The total width is just the sum of the partial widths: $\Gamma = \sum_i\Gamma_i$. For a given resonance, the values of $\Gamma$ and $\Gamma_i$ listed in Tables I and II were evaluated at the energy $W=M$. The first column also gives the PDG star rating \cite{pdg12} of each resonance. The second column lists the fitted  decay channels, starting with the $\overline K N$ elastic channel. 
Sometimes subscripts appear with a channel notation (e.g. $(\overline K^*_3 N)_D$); here the first subscript is twice the total intrinsic spin (2S) and the second subscript denotes the orbital angular momentum of the channel. The third column in Table I or II lists the partial decay widths ($\Gamma_i$) associated with the corresponding channels. The symbol ${\cal B}_i$ in the fourth column denotes the branching ratio for a given channel. Finally, the $x$ and $x_i$ represent the ratio of elastic partial width and partial width for the $i^{th}$ channel respectively to the total width. 
Any resonance included above 2.1 GeV had its mass parameter initially fixed and resonance parameters for these states are generally not listed. The uncertainties of the resonance parameters in Tables I and II were conservatively increased by a factor of $\sqrt{\chi^2/\nu}$  if $\chi^2/\nu> 1$, where $\chi^2/\nu$ was the $\chi^2$ per degree of freedom for the fit.

Figures 1 - 3 show  representative Argand diagrams for $I=0$ partial waves ($\overline KN, \pi\Sigma$) and Figs. 4 - 7 show representative Argand diagrams for $I = 1$ partial waves ($\overline K N, \pi\Lambda$, and $ \pi\Sigma$). 
The Argand diagrams display the energy dependence of the partial-wave $T$-matrix amplitudes. The relationship between the $T$-matrix and the unitary partial-wave $S$-matrix is $S=I + 2{\rm i}T$, where $I$ is an identity matrix. For the case of a single resonance in the absence of background, a $T$-matrix amplitude traces out a counter-clockwise circle on the Argand diagram. For that simple case, the Breit-Wigner mass of the resonance corresponds to the c.m.\ energy in which the real part of the amplitude vanishes and the magnitude of the imaginary part is maximum. For elastic ($\overline KN\rightarrow \overline KN$) amplitudes, the value of the imaginary part of the amplitude at resonance corresponds to the elasticity, $x$, of the resonance (here, the $\overline KN$ branching fraction). More generally, for a single resonance in the absence of background, the value of imaginary part of the $T$-matrix amplitude at resonance is $\sqrt{xx_i}$. For inelastic amplitudes, the sign of $\sqrt{xx_i}$ is positive or negative depending on the sign of the amplitude at resonance. For more realistic situations in which there are two or more resonances and non-resonant background, the appearance of Argand diagrams becomes more complex and one must rely on fits in order to be able to extract resonance parameters. In the following, we discuss the results of our multichannel fits that take these considerations into account.  
To discuss the resonance parameters we follow a logical sequence of partial waves.\\

\begin{figure*}[htpb]
\centering
\vspace{-21mm}
\centerline{\includegraphics[scale=1.0]{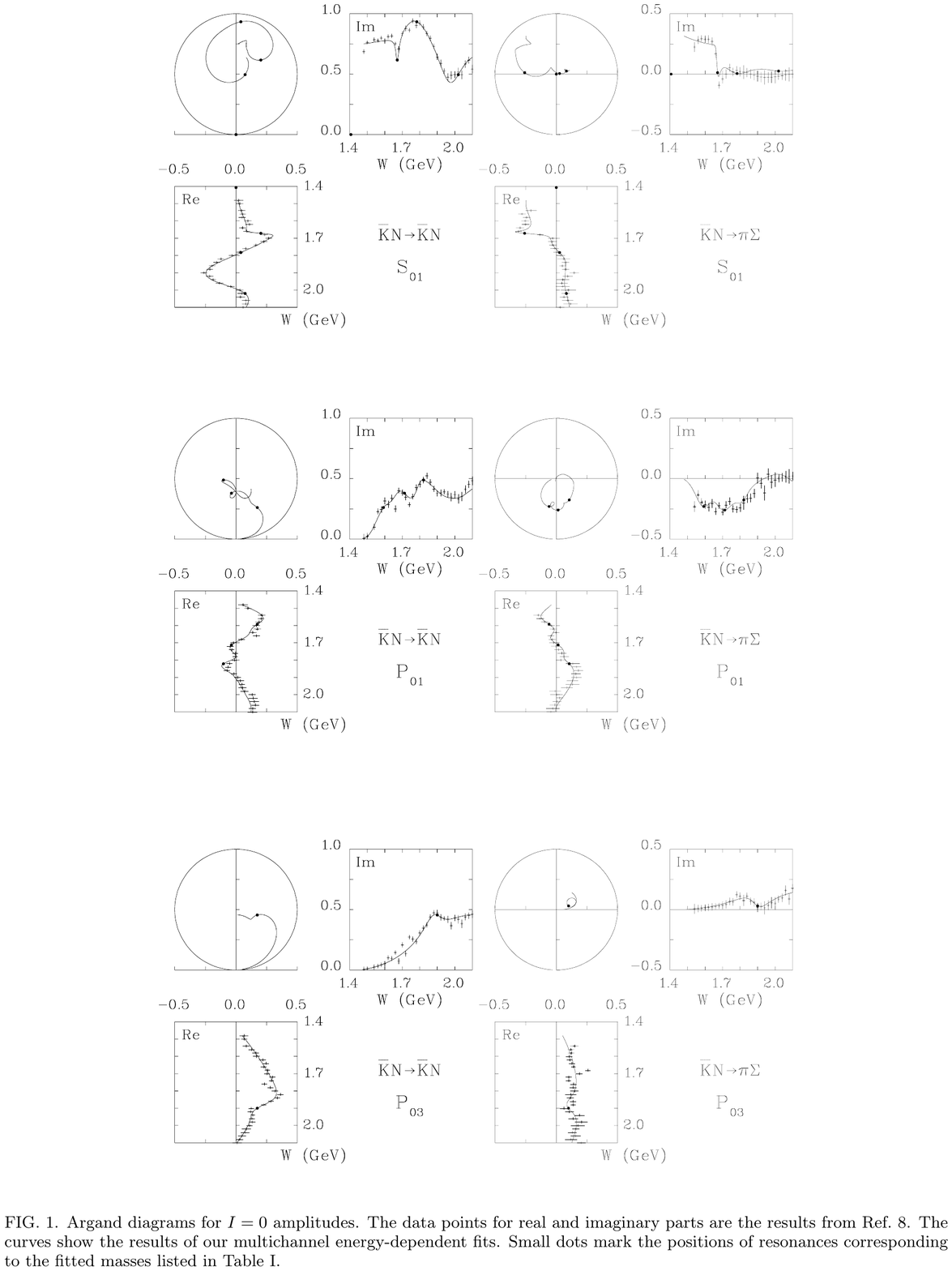}}
\vspace{-40mm}
\end{figure*}

\begin{figure*}[htpb]
\centering
\vspace{-21mm}
\centerline{\includegraphics[scale=1.0]{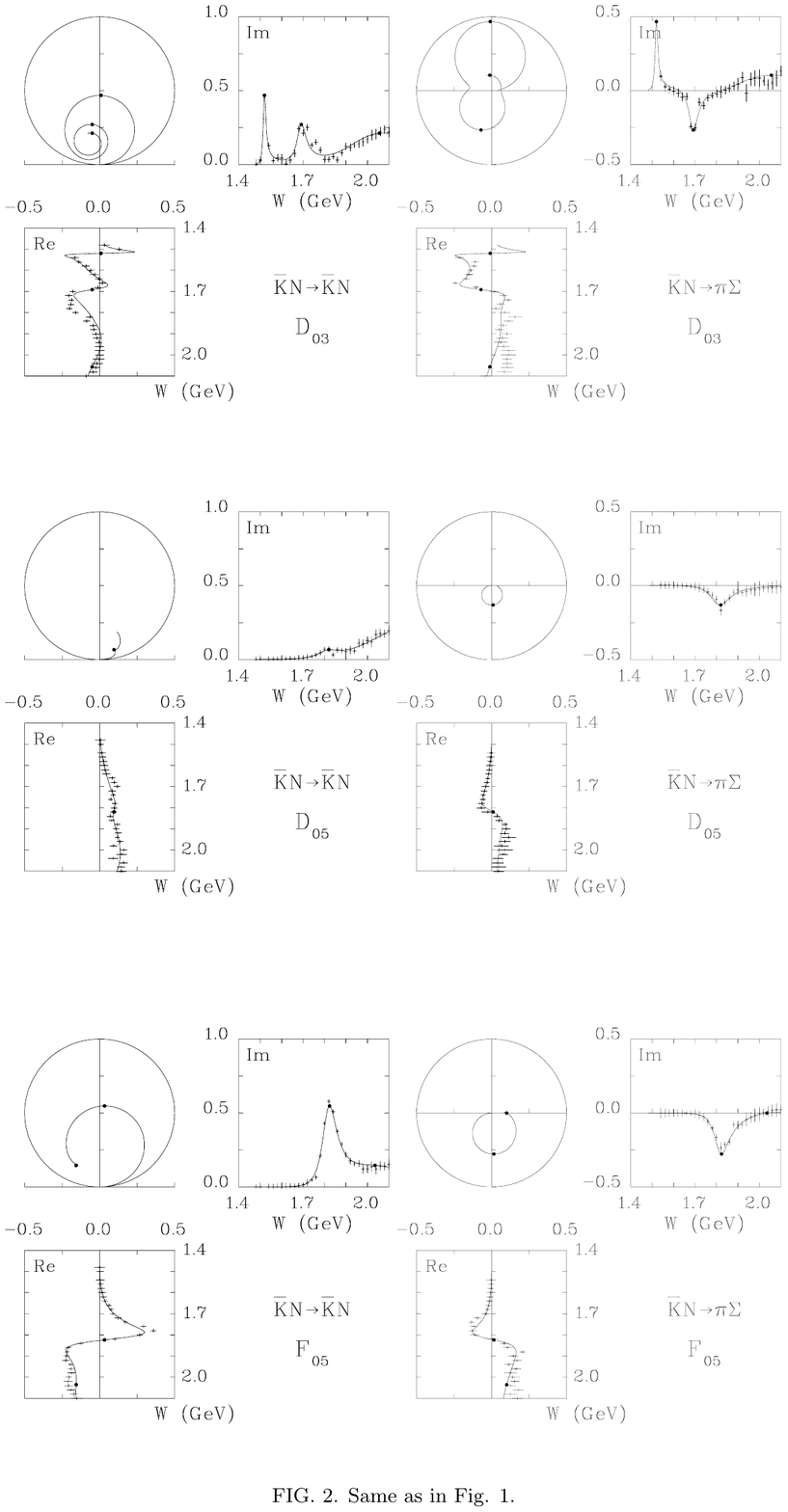}}
\vspace{-40mm}
\end{figure*}

\begin{figure*}[htpb]
\centering
\vspace{-21mm}
\centerline{\includegraphics[scale=1.0]{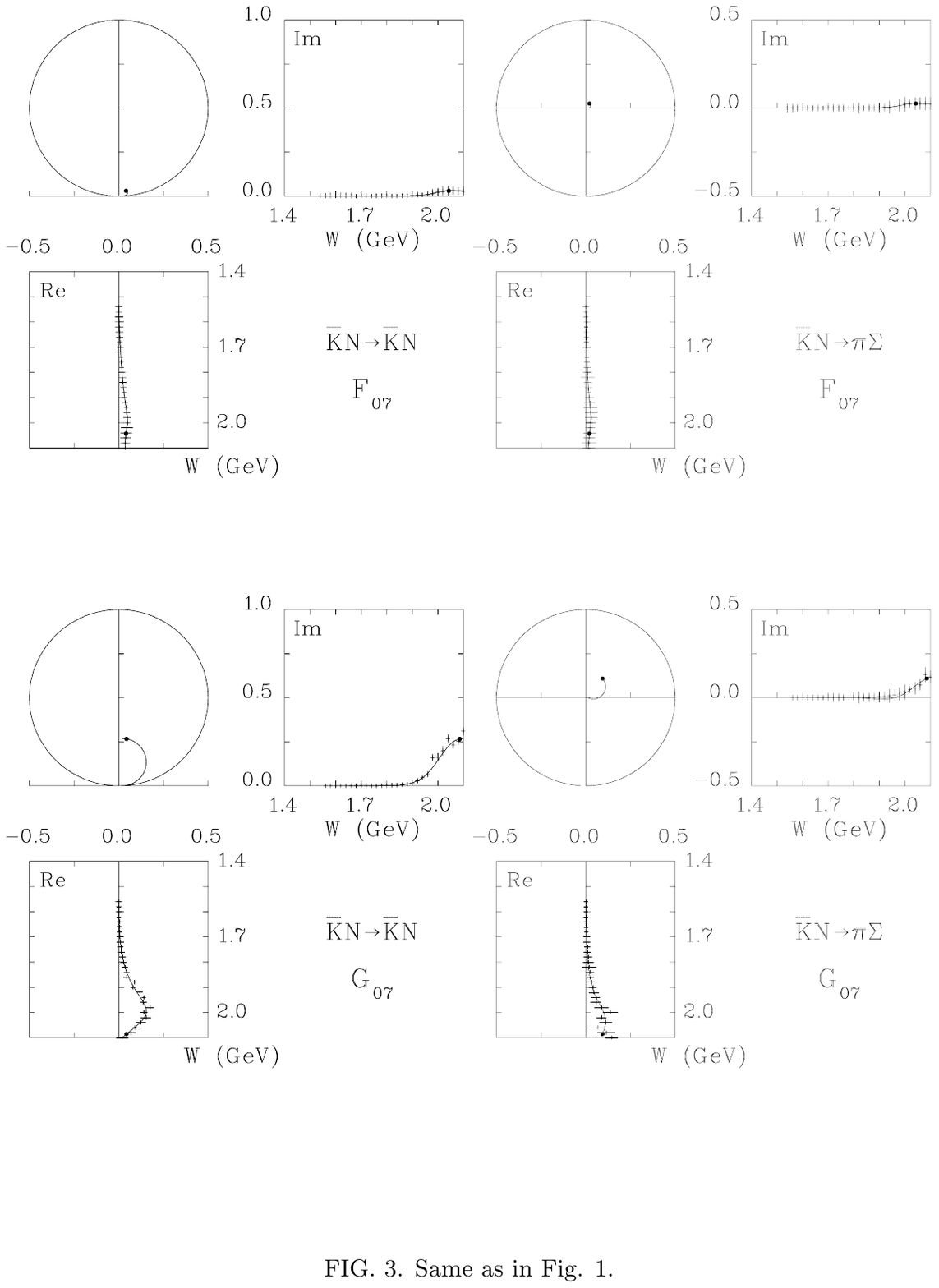}}
\vspace{-40mm}
\end{figure*}

\begin{figure*}[htpb]
\centering
\vspace{-21mm}
\centerline{\includegraphics[scale=1.0]{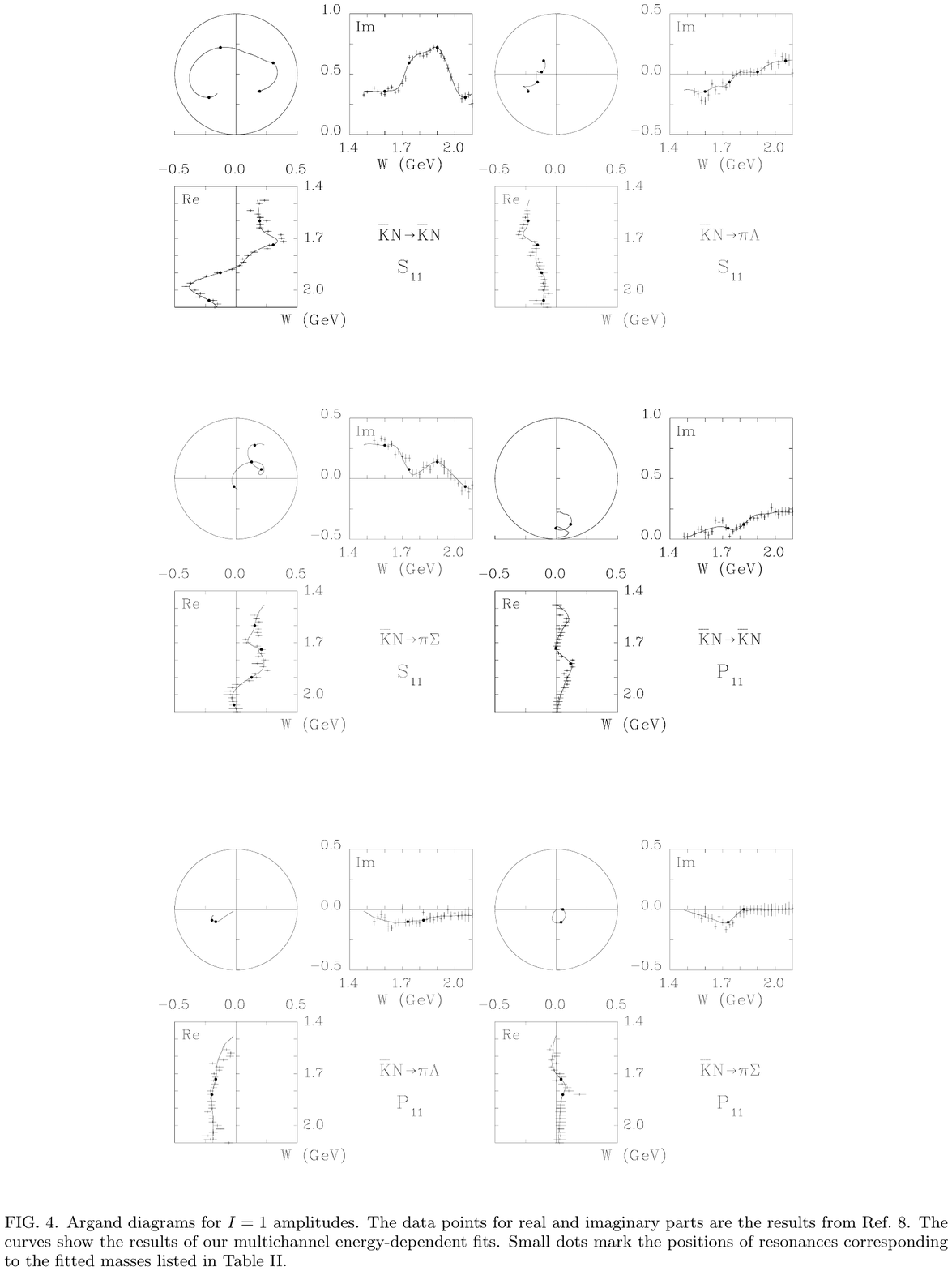}}
\vspace{-40mm}
\end{figure*}

\begin{figure*}[htpb]
\centering
\vspace{-21mm}
\centerline{\includegraphics[scale=1.0]{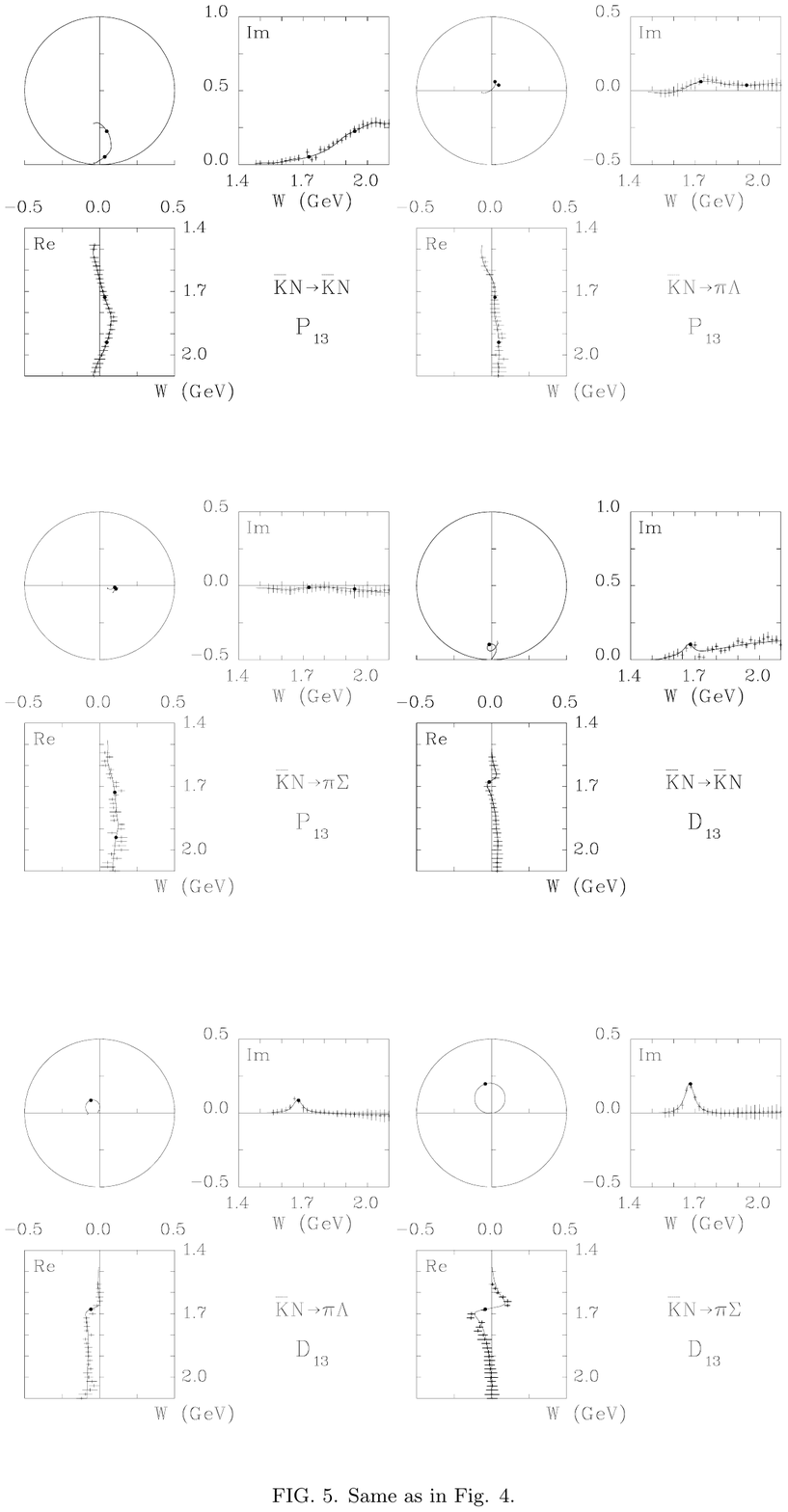}}
\vspace{-40mm}
\end{figure*}

\begin{figure*}[htpb]
\centering
\vspace{-21mm}
\centerline{\includegraphics[scale=1.0]{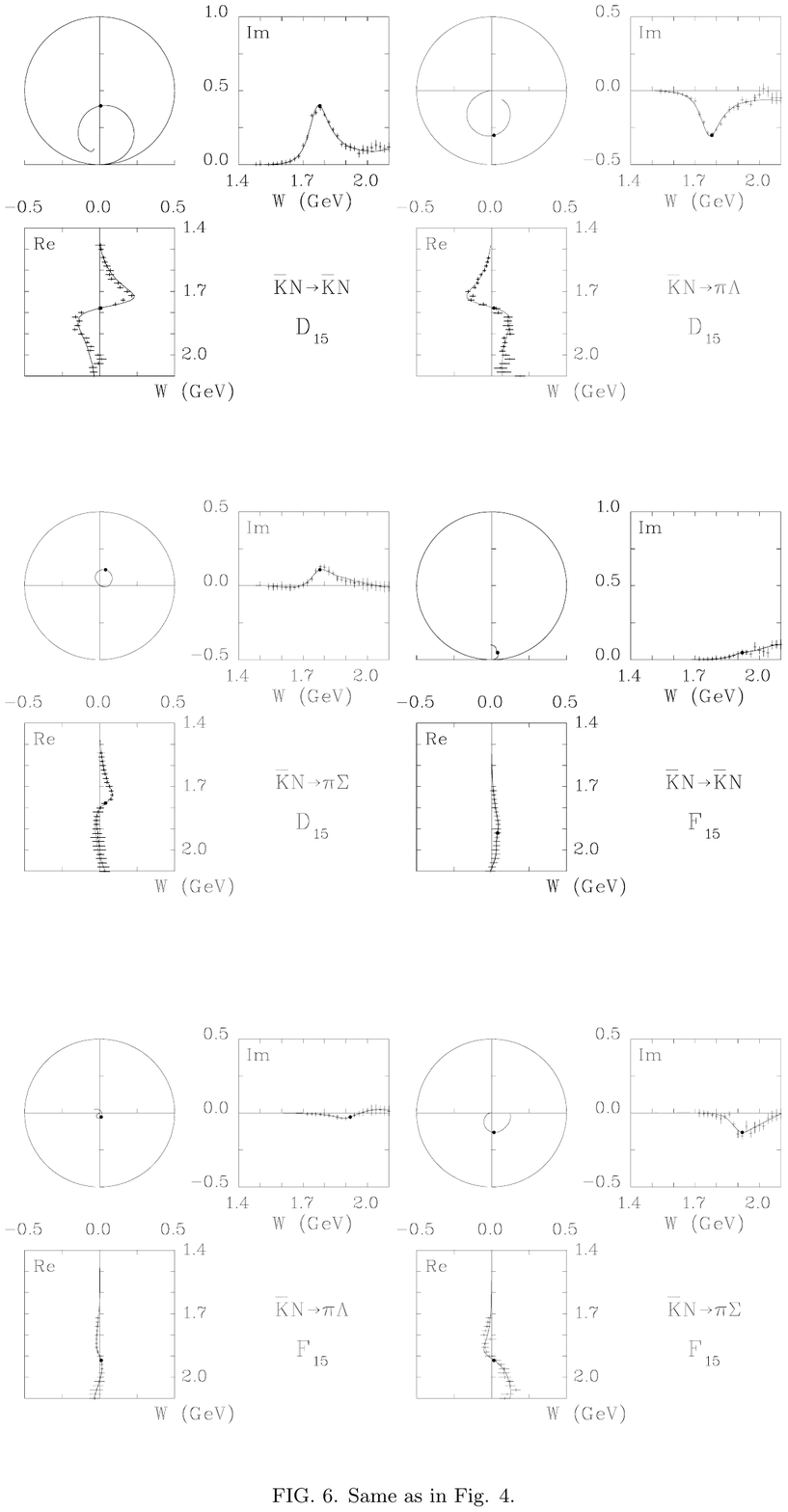}}
\vspace{-20mm}
\end{figure*}

\begin{figure*}[htpb]
\centering
\vspace{-21mm}
\centerline{\includegraphics[scale=1.0]{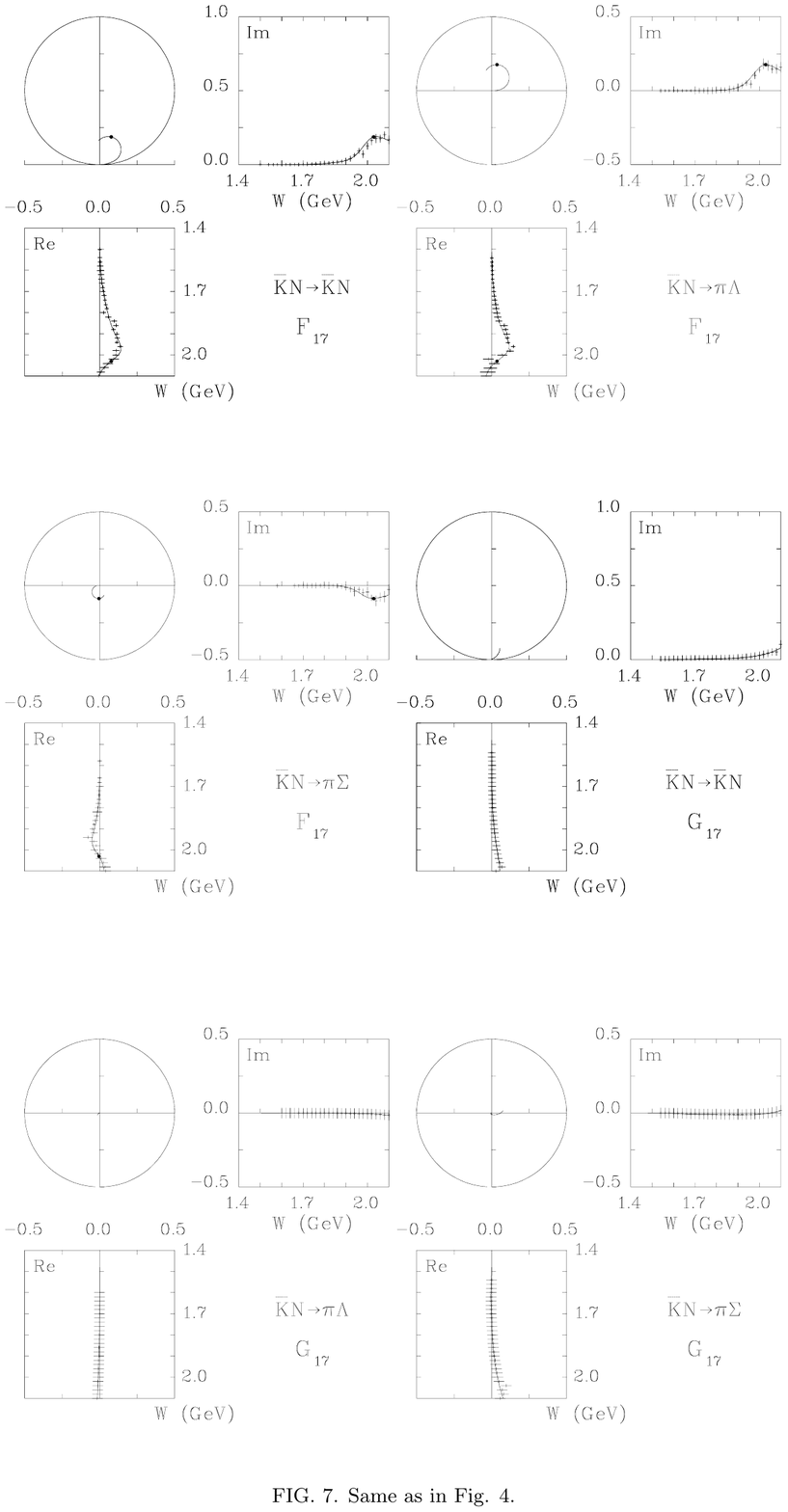}}
\vspace{-40mm}
\end{figure*}



$\bf{S_{01}}$:\\
This partial wave was fitted with four resonances. Among which, the first resonance $\Lambda (1405){S_{01}}$ lies below $\overline{K} N$ threshold and nearly everything about it was learned from production experiments {\cite{pdg12}}. Therefore, in our fits, all of its parameters (mass $M$ = 1406.50 MeV, width $\Gamma = 49.98$ MeV, $\pi\Sigma$ branching ratio = 100\%) were held fixed. The second resonance occurred with a mass $M = 1672\pm 3$ MeV and width $\Gamma$ = $29\pm5$ MeV and corresponds to the 4* $\Lambda (1670)S_{01}$. The strength of this resonance divides more or less equally to $\overline K N$, $\pi \Sigma$, and $\eta \Lambda$ at 26\%, 31\%, and 34\%, respectively. 
The third resonance was found to have $M =  1783\pm19$ MeV and $\Gamma$ = $256\pm35$ MeV corresponding to the 3* $\Lambda(1800)S_{01}$. Decay modes for this state are primarily $\overline K N$, $\eta\Lambda$, $\sigma\Lambda$(dummy), $\overline K^*_1 N$, and $\pi\Sigma$. We found the fourth resonance with $M = 2020\pm16$ MeV and $\Gamma = 255\pm63$ MeV corresponding to the 1* $\Lambda(2000)S_{01}$. 


$\bf{P_{01}}$:\\
This partial wave was fitted with four resonances. The first resonance occurred at $M = 1592\pm10$ MeV with $\Gamma = 150\pm 28$ MeV and corresponds to the 3* $\Lambda(1600) P_{01}$. 
The main decay modes are $\overline K N$, $\pi\Sigma$, and $\sigma\Lambda$ (dummy) channels.
Our analysis suggests the existence of the state $\Lambda(1710)P_{01}$, which is compatible with the proposed $\Lambda(1740)$ state in Ref.  \cite{Isgur79}. This resonance occurred at $ M=1713\pm 13$ MeV with $\Gamma = 180\pm42$ MeV. Its elasticity is about 43\%. 
The third resonance occurred at $ M=1821\pm10$ MeV with $\Gamma = 174\pm50$ MeV. This resonance corresponds to the 3* $\Lambda(1810)P_{01}$. The major decay modes are $\overline K N$ and $(\overline K^*_3N)_P$. The fourth resonance was found at $ M=2151\pm27$ MeV with $\Gamma=300\pm81$ MeV. 


$\bf{P_{03}}$:\\
This partial wave was fitted with two resonances. The first resonance occurred at $ M=1900\pm5$ MeV with $\Gamma = 161\pm15$ MeV and can be identified with the 4* $\Lambda(1890)P_{03}$. 
The major decay modes were found to be $\overline K N$ (37\%), ($\pi\Sigma^*(1385))_F(27\%)$, and $\sigma\Lambda$ (23\%) (dummy channel). 
A second resonance was included at $ M = 2284$ MeV with a width broader than 500 MeV.


$\bf{D_{03}}$:\\
This partial wave was fitted with three resonances. The first resonance occurred at $ M=1519.6\pm 0.5$ MeV with $\Gamma= 17\pm1$ MeV, and corresponds to the 4* $\Lambda(1520)D_{03}$. Our results for this state agree very well with those from prior analyses. 
This state was found to decay equally via elastic and $\pi\Sigma$ channels at 47\% each. The second resonance occurred at $ M=1691\pm3$ MeV with $\Gamma=54\pm5$ MeV. The major decay modes were $\overline K N$, $(\pi\Sigma^*(1385))_S$, and $\pi\Sigma$. The third resonance occurred at $ M=2056\pm22$ MeV with $\Gamma=493\pm61$ MeV. Its major decay channel was a dummy $\sigma\Lambda$ (40\%) channel and its elasticity was found to be about 19\%. 

      
$\bf{D_{05}}$:\\
Two resonances were required to fit this partial wave. The first resonance occurred at $ M=1820\pm4$ MeV with $\Gamma = 114\pm10$ MeV. It corresponds to the 4* $\Lambda(1830)D_{05}$. Our results for this state agree very well with those from previous analyses. The major decay channels were $(\pi\Sigma^*(1385))_D$ (52\%),  $\pi\Sigma$ (42\%), and $\overline K N$ (4\%). A second resonance was included at $M = 2292$ MeV with a width more than 500 MeV.


$\bf{F_{05}}$:\\
This partial wave was fitted with two resonances. The first resonance occurred at  $ M= 1823.5\pm0.8$ MeV with $\Gamma=89\pm2$ MeV and can be identified with the 4* $\Lambda(1820)F_{05}$. This resonance was found to be highly elastic with an elasticity of 54\%. The elastic amplitude for this state exhibited classic Breit-Wigner behavior. The other major channels were $(\pi\Sigma^*(1385))_P$ (8\%), $\pi\Sigma$ (15\%), $(\overline K^*_3N)_P$ (4\%), and $\sigma\Lambda$ ($20\%$) (dummy channel). The second resonance occurred at $ M=2036\pm13$ MeV with $\Gamma=400\pm38$ MeV and probably corresponds to the 3* $\Lambda(2110)F_{05}$ although our mass is somewhat lower than the range given by the PDG \cite{pdg12}. 
Its major decay modes were $\overline K N$ (8\%), $\overline K^*_1 N$ (10\%), $(\overline K^*_3N)_P$ (67\%), and $\sigma\Lambda$ (11\%) (dummy channel).


$\bf{F_{07}}$:\\
This partial wave was fitted with a single resonance at $ M=2043\pm22$ MeV and $\Gamma=200\pm75$ MeV. This resonance is highly inelastic with an elasticity of only 3\%. Its major decay modes were $\overline K_1^*N$ (30\%) and $\sigma\Lambda$ ($65\%$) (dummy channel).


$\bf{G_{07}}$:\\
This partial wave was fitted with one resonance at $ M=2086\pm6$ MeV with $\Gamma=305\pm16$ MeV. This resonance corresponds to the 4* $\Lambda(2100)G_{07}$. Its major decay channels were $\overline K N$ (23\%), $(\overline K^*_3N)_D$ (11\%), $(\overline K^*_3N)_G$ (3\%), and $\sigma\Lambda$ ($63\%$) (dummy channel). 


$\bf{S_{11}}$:\\
Four resonances were required to fit this partial wave. The first resonance occurred at $ M=1600\pm15$ MeV with $\Gamma=400\pm152$ MeV and can be identified with the 2* $\Sigma(1620) S_{11}$. The primary decay modes for this state were found to be $\overline K N$ (59\%) and $\pi \Sigma$ (17\%). The second resonance occurred at $M=1739\pm 8$ MeV with $\Gamma = 182\pm60$ MeV and corresponds to the 3* $\Sigma(1750)S_{11}$. Its major decay modes were found to be $\pi\Lambda$ (12\%), $\pi\Sigma$ (30\%), and $(\pi\Sigma^*(1385))_D$ (33\%). The third resonance occurred at $M=1900\pm21$ MeV with $\Gamma= 191\pm47$ MeV. This state was found to be highly elastic with an elasticity of 67\%. The fourth resonance was found at $ M=2060\pm20$ MeV with $\Gamma=300\pm134$ MeV. This state can be identified with the 1* $\Sigma(2000)S_{11}$. It was found to be highly inelastic with the dominant decay mode a dummy $\eta\Sigma$ (71\%) channel.


$\bf{P_{11}}$:\\
This partial wave was fitted with two resonances. The first resonance occurred at $ M=1732\pm28$ MeV with $\Gamma=200\pm 41$ MeV. The dominant decay channels were found to be $\pi \Sigma$ (32\%) and $(\pi\Sigma^*(1385))_P$ (42\%). This resonance can be identified with the 1* $\Sigma(1770)P_{11}$. The second resonance occurred at $ M = 1821\pm17$ MeV with $\Gamma = 300\pm59$ MeV and corresponds to the 2* $\Sigma(1880)P_{11}$. Its major decay modes were found to be $\overline K N$ (10\%), $(\overline K\Delta)_P$ (39\%), and $\sigma\Sigma$ (47\%) (dummy channel).


$\bf{P_{13}}$:\\
This partial wave was fitted with two resonances. The first resonance occurred at $ M=1727\pm27$ MeV with $\Gamma=276\pm87$ MeV and  marks a new state $\Sigma(1730)P_{13}$. This state has an elasticity of just 2\%. The dominant decay channels are $\pi\Sigma$ (12\%) and $\pi\Lambda$ (70\%). The second resonance occurred at $M=1941\pm18$ MeV with $\Gamma=400\pm49$ MeV and corresponds to another new state. Its major decay modes were found to be $\overline K N$ (13\%), $(\pi\Sigma^*(1385))_P$  (22\%), and $\sigma\Sigma$ (52\%) (dummy channel). 


$\bf{D_{13}}$:\\
This partial wave was fitted with one resonance at $ M=1678\pm2$ MeV and $\Gamma=55\pm4$ MeV. This state can be identified with the 4* $\Sigma(1670)D_{13}$. Its main decay modes were found to be $\overline K N$ (6\%), $\pi\Sigma$ (62\%), and $\pi\Lambda$ (11\%). 


$\bf{D_{15}}$:\\
This partial wave was fitted with two resonances. The first resonance occurred at $ M=1778\pm1$ MeV with $\Gamma=131\pm3$ MeV. This resonance can be identified with the 4* $\Sigma(1770)D_{15} $. This state was found to have an elasticity of 40\% with the main inelasticity due to $\pi\Lambda$ (25\%) and $\sigma\Sigma$ (28\%) (dummy channel). A second resonance was included at $M = 2201\pm16$ MeV with $\Gamma = 300\pm43$ MeV.


$\bf{F_{15}}$:\\
This partial wave was fitted with two resonances. The first resonance occurred at $ M=1920\pm7$ MeV with $\Gamma=149\pm17$ MeV. This resonance can be identified with the 4* $\Sigma(1915)F_{15}$. Its major decay channels were found to be $\overline K N$ (3\%), $\pi\Sigma$ (76\%), $(\pi\Sigma^*(1385))_F$  (13\%), and $(\overline K_3^*N)_F$ (5\%). The second resonance occurred at $ M=2124\pm13$ MeV with $\Gamma=339\pm42$ MeV. 


$\bf{F_{17}}$:\\
This partial wave was fitted with one resonance at $ M=2030\pm5$ MeV with $\Gamma=207\pm17$ MeV. This state can be identified with the 4* $\Sigma(2030)F_{17}$. This state was found to have an elasticity of 13\% with the main inelasticity due to $\pi\Lambda$ (18\%), $(\pi\Sigma^*(1385))_F$ (19\%), $(\overline K\Delta)_F$ (12\%), and  $\sigma\Sigma$ (29\%) (dummy channel), respectively.


$\bf{G_{17}}$:\\
This partial wave was fitted with one resonance at $ M=2241\pm40$ MeV with $\Gamma=292\pm91$ MeV.\\

\begin{table*}[htpb]
\caption {Resonance parameters for states with isospin $I = 0$.  Column 1 lists the resonance name
followed by its fitted mass (in MeV), fitted total width (in MeV), and star rating. Column 2 lists the decay channel (see text for explanation). Column 3 lists the partial width in MeV and column 4 lists the corresponding branching fraction. Column 5 lists the resonant amplitude (see text). Contributions from dummy channels are not listed (see text).\label{Resonances} }
\begin{center}
\begin{ruledtabular}
\begin{tabular}{cccccccccc}

{Resonance}    & {Channel} & {$\Gamma_i$ (MeV)} & {${\cal B}_i$ ($\%$)}  & {$\sqrt{xx_i}$ } &{Resonance}    & {Channel} & {$\Gamma_i$ (MeV)} & {${\cal B}_i$ ($\%$)}  & {$\sqrt{xx_i}$ }\\ \hline
      $\Lambda(1670)S_{01}$  &  $\overline{K} N$   & $<16$  & $26(25)$  & $+0.26(25)$    &       $\Lambda(1800)S_{01}$  &  $\overline{K} N$   & $34(14)$  & $13(6)$  & $+0.13(6)$\\
      $1672(3)$  & $\pi \Sigma$  &  $9(6)$  &  $31(22)$ &  $-0.29(6)$      &                                     $1783(19)$  & $\pi \Sigma$   &  $10(7)$  &  $4(3)$ &  $-0.07(2)$\\
      $29 (5)$  & $(\pi \Sigma^*(1385))_D$ & $<2$ & $<5$ & $-0.06(10)$                        &      $256(35)$  &  $(\pi \Sigma^*(1385))_D$ & $15(14)$ & $6(5)$ & $-0.09(5)$\\
      $\ast\ast$$\ast\ast$& $\overline{K}^*_1 N$  &  $<1$  &  $<2$  &  $+0.02(11)$             & $\ast\ast$$\ast$             & $\overline{K}^*_1 N$  &  $33(11)$  &  $13(4)$  &  $-0.13(2)$\\
      &$(\overline{K}^*_3 N)_D$ &$2(1)$&$5(4)$& $-0.12(6)$                                                  &       & $(\overline{K}^*_3 N)_D$   &  $<2$  & $<1$ &   $+0.02(2)$\\
      &$\eta \Lambda$ &$10(3)$	&	$34(11)$	&	$-0.30(10)$ 		     &     &    $\eta \Lambda$ &$14(14)$	&	$6(5)$	&	$+0.09(5)$  \\
    & &  & & &                                                                                                             & &&&\\
      $\Lambda(2000)S_{01}$   & $\overline{K} N$     & $69(16)$      & $27(6)$ &   $+0.27(6)$       &    $\Lambda(1600)P_{01}$  &  $\overline{K} N$   & $21(8)$  & $14(4)$  & $+0.14(4)$\\
      $2020(16)$ & $\pi \Sigma$   & $4(4)$      & $2(2)$        &   $-0.07(3)$                                             &     $1592(10)$  & $\pi \Sigma$   &  $57(17)$  &  $38(10)$ &  $-0.23(3)$\\
      $255(63)$    & $(\pi \Sigma^*(1385))_D$   &$ <3$        &$<2$          &   $-0.02(5)$                              &     $150(28)$  & $(\pi \Sigma^*(1385))_P$ & $<9$ & $<6$ & $-0.04(9)$\\
        $(New)$                     & $\overline{K}^*_1 N$           &  $<12$    & $<5$      & $-0.06(5)$ &                             $\ast\ast$$\ast$&                                $\overline{K}^*_1 N$ &  $1(1)$  &  $<1$  &  $-0.03(1)$\\
                             & $(\overline{K}^*_3 N)_D$          & $110(44)$          & $43(11)$   & $+0.34(5)$&             & $(\overline{K}^*_3 N)_P$  &$<1$& $<1$& $-0.01(2)$\\
                           &$\eta \Lambda$ &$41(17)$	&	$16(7)$	&	$-0.21(5)$ 		     &     &    &&& \\
     & &  & & &                                                                                                             & &&&\\
       $\Lambda(1710)P_{01}$  &  $\overline{K} N$   & $78(21)$    & $43(4)$&  $+0.43(4)$&               $\Lambda(1810)P_{01}$ &  $\overline{K} N$   & $34(11)$  & $19(8)$  & $+0.19(8)$\\
      $1713(13)$  &$\pi \Sigma$  &  $38(16)$           & $21(5)$          &  $-0.30(4)$&                $1821(10)$  & $\pi \Sigma$   &  $6(5)$  &  $3(3)$&  $-0.08(5)$\\
      $180(42)$  & $(\pi \Sigma^*(1385))_P$     & $36(17)$       & $20(8)$           &  $+0.29(6)$&      $174(50)$  &   $(\pi \Sigma^*(1385))_P$  & $<1$ & $<1$ & $+0.01(4)$\\
       $(New)$& $\overline{K}^*_1 N$  &  $9(8)$    & $5(4)$       & $+0.15(6)$&                                    $\ast$& $\overline{K}^*_1 N$  &  $<4$  &  $<3$  &  $+0.03(5)$\\
       & $(\overline{K}^*_3 N)_P$  &  $18(13)$    & $10(8)$     & $-0.21(8)$&                                            & $(\overline{K}^*_3 N)_P$  &  $130(50)$  &  $75(10)$  &  $+0.38(6)$\\
  & &  & & &                                                                                                             & &&&\\
       $\Lambda(1890)P_{03}$  & $\overline{K} N$   & $59(8)$  & $37(3)$  & $+0.37(3)$ & $\Lambda(1520)D_{03}$  &  $\overline{K} N$   & $7.7(4)$  & $47(4)$  & $+0.47(4)$\\
        $1900(5)$  & $\pi \Sigma$  &  $3(2)$  & $2(1)$  &  $-0.09(2)$   &           $1519.6(5)$  & $\pi \Sigma$   &  $8(1)$  &  $47(5)$ &  $+0.47(3)$\\
        $161(15)$  &  $(\pi \Sigma^*(1385))_P$   & $<4$    &   $<3$ & $-0.06(4)$   & $17(1)$  &  $(\pi \Sigma^*(1385))_S$ & $<3$ & $<13$ & $-0.12(19)$\\
        $\ast\ast$$\ast\ast$&$(\pi \Sigma^*(1385))_F$   & $43(9)$    &   $27(6)$ & $-0.31(4)$  &         $\ast\ast$$\ast\ast$& $(\pi \Sigma^*(1385))_D$  &  $<1$  &  $<3$  &  $-0.04(11)$\\
     & $\overline{K}^*_1 N$ &  $12(7)$       &  $8(4)$   &    $-0.17(5)$ &&&&&\\
     & $(\overline{K}^*_3 N)_P$   &  $<1$         &  $<1$      &      $-0.02(5)$&&&&&\\
     & $(\overline{K}^*_3 N)_F$   &  $5(3)$         &  $3(2)$      &      $-0.11(3)$ &&&&& \\                                   
   
  & &  & & &                                                                                                             & &&&\\
          $\Lambda(1690)D_{03}$  &  $\overline{K} N$   & $13(2)$  & $25(4)$  & $+0.25(4)$  & $\Lambda(2050)D_{03}$   &  $\overline{K} N$  & $96(24)$  & $19(4)$  & $+0.19(4)$\\
                   $1691(3)$  & $\pi \Sigma$   &  $16(4)$  &  $30(7)$ &  $-0.27(3)$                        & $2056(22)$  & $\pi \Sigma$   &  $30(18)$  &  $6(3)$ &  $+0.11(3)$\\ 
                  $54(5)$  &  $(\pi \Sigma^*(1385))_S$ & $17(6)$ & $31(11)$ & $-0.28(6)$            &$493(61)$  &  $(\pi \Sigma^*(1385))_S$ & $41(26)$   & $8(6)$ & $-0.13(4)$\\
           $\ast\ast$$\ast\ast$      & $(\pi \Sigma^*(1385))_D$  &  $<5$  &  $<9$  &  $-0.05(18)$ &($New$) & $(\pi \Sigma^*(1385))_D$      &  $19(17)$       &  $4(3)$    &  $-0.09(4)$\\
                & $\overline{K}^*_1 N$   &  $<4$  & $<6$ &  $-0.08(5)$                                                   && $\overline{K}^*_1 N$     &  $111(38)$         & $23(7)$        &   $-0.21(3)$\\
      & &  & & &                                                                                                             & &&&\\

            $\Lambda(1830)D_{05}$  &  $\overline{K} N$   & $4.7(6)$  & $4.1(5)$  & $+0.04(1)$ &&&&&\\
                                                $1820(4)$      &  $\pi \Sigma$   &  $48(7)$  &  $42(5)$ &  $-0.13(1)$ &&&&&\\
                             $114(10)$      &  $(\pi \Sigma^*(1385))_D$ & $59(9)$ & $52(6)$ & $+0.15(1)$ &&&&&\\
                           $\ast\ast$$\ast\ast$              & $(\pi \Sigma^*(1385))_G$   &  $<5$  &  $<5$  &     $-0.03(2)$ &&&&&\\

    
    \end{tabular}
    \end{ruledtabular}
    \end{center}
    \end{table*}
   
    \begin{table*}[htbp]
    \addtocounter{table}{-1}
    \caption{Cont'd.}
       \begin{center}
      \begin{ruledtabular}
\begin{tabular}{cccccccccc}
{Resonance}    & {Channel} & {$\Gamma_i$ (MeV)} & {${\cal B}_i$ ($\%$)}  & {$\sqrt{xx_i}$ } &{Resonance}    & {Channel} & {$\Gamma_i$ (MeV)} & {${\cal B}_i$ ($\%$)}  & {$\sqrt{xx_i}$ }\\  \hline
        $\Lambda(1820)F_{05}$ &  $\overline{K} N$   & $48(1)$  & $54(1)$  & $+0.54(1)$&    $\Lambda(2110)F_{05}$  &  $\overline{K} N$   & $33(3)$  & $8.3(5)$  & $+0.08(1)$\\
      $1823.5(8)$& $\pi \Sigma$  &  $13(1)$  &  $15(1)$  &  $-0.28(1)$&                              $2036(13)$  & $\pi \Sigma$   &  $8(3)$  &  $2.0(7)$ &  $+0.04(1)$\\
      $89(2)$ & $(\pi \Sigma^*(1385))_P$   &  $7(1)$  & $8(1)$ &   $-0.20(2)$&                       $400(38)$   &  $(\pi \Sigma^*(1385))_P$ & $7(4)$ & $2(1)$ & $+0.04(1)$\\
 $\ast\ast$$\ast\ast$     & $(\pi \Sigma^*(1385))_F$    &  $<1$  & $<1$ &  $+0.05(1)$&           $\ast\ast$$\ast$                                                      & $(\pi \Sigma^*(1385))_F$    &  $<4$  & $<1$ &  $-0.02(1)$\\
      & $\overline{K}^*_1 N$    & $<1$  &$ <1$ & $+0.01(1)$&                                                                             &$\overline{K}^*_1 N$   &  $39(11)$  & $10(2)$ &  $-0.09(1)$\\
      &$(\overline{K}^*_3 N)_P$    & $3(1)$  & $3(1)$  & $+0.14(1)$&                                                              &$(\overline{K}^*_3 N)_P$&$268(33)$&$67(6)$&$+0.24(1)$\\
      & $(\overline{K}^*_3 N)_F$    & $<1$  & $<1$  & $+0.00(1)$&                                                             & $(\overline{K}^*_3 N)_F$    & $<1$  & $<1$  & $+0.01(1)$\\
   & &  & & &                                                                                                             & &&&\\   
     $\Lambda(2020)F_{07}$  &  $\overline{K} N$    & $6(2)$  & $2.8(5)$  & $+0.03(1)$         &               $\Lambda(2100)G_{07}$  &  $\overline{K} N$    & $70(6)$  & $23(1)$  & $+0.23(1)$\\
      $2043(22)$  & $\pi \Sigma$   &  $4(2)$  &  $2(1)$ &  $+0.02(1)$     &                                                   $2086(6)$  &       $\pi \Sigma$   &  $1(1)$  &  $<1$ &  $+0.03(1)$\\
     $200(75)$  & $\overline{K}^*_1 N$  &$60(26)$   & $30(9)$  &      $-0.09(1)$ &                                                        $305(16)$  &  $\overline{K}^*_1 N$  & $1(1)$ & $1(1)$ & $-0.03(2)$\\
       $\ast$                     &  &&  &                                                     &                                                                       $\ast\ast$$\ast\ast$     & $(\overline{K}^*_3 N)_D$   &  $33(7)$  &  $11(2)$  &  $+0.16(2)$\\
      &    &   &    &            &                                                                                                                                                                  &$(\overline{K}^*_3 N)_G$     & $8(4)$     & $3(1)$    &   $+0.08(2)$\\
    \end{tabular}
    \end{ruledtabular}
    \end{center}
    \end{table*}

\begin{table*}[htpb]
 \caption {Resonance parameters for states with isospin $I = 1$. (See caption to Table I for details.)}
\begin{center}
\begin{ruledtabular}
 \begin{tabular}{cccccccccc}

  Resonance     &Channel & $\Gamma_i$ (MeV) & ${\cal B}_i$ ($\%$) & $\sqrt{xx_i}$ &{Resonance}    & {Channel} & {$\Gamma_i$ (MeV)} & {${\cal B}_i$ ($\%$)}  & {$\sqrt{xx_i}$ }\\ \hline
 
      $\Sigma(1620) S_{11}$  &$\overline{K} N$ 	&	237(117)	& 59(10)	& $+0.59(10)$  &    $\Sigma(1750) S_{11}$ &  $\overline{K} N$ 	&	$<34$	&	$9(7)$	&	$+0.09(7)$	 \\           
      $1600(15)$  & $\pi \Lambda$ 	&	$<38$	&	$<9$	&	$+0.16(9)$ &               $1739(8)$   & $\pi \Lambda$ 	&	$22(10)$	&	$12(4)$	&	$+0.10(4)$\\
      $400(152)$  &$\pi \Sigma$ 	&	$69(23)$	&	$17(3)$	&	$+0.32(3)$	 &                             $ 182(60)$&  $\pi \Sigma$ 	&	55(16)	&	30(5)	&	+0.17(7)     \\
      $\ast\ast$&$(\pi \Sigma^*(1385))_D$	&	$<30$	&	$<8$	&	$+0.14(8)$	              &   $\ast$$\ast\ast$ &          $(\pi \Sigma^*(1385))_D$	&	$60(29)$&	$33(11)$	&	 $+0.17(7)$	\\                                                                                                        
      &$(\pi \Lambda^*(1520))_P$	&	$<5$	&	$<2$	&	$-0.06(3)$	 && $(\pi \Lambda^*(1520))_P$	&	$<14$		&	$<8$	&	$-0.04(6)$\\
      & $(\overline{K} \Delta)_D$	&	$<8$	&	$<2$	&	$+0.07(4)$	& & $(\overline{K} \Delta)_D$	&	$<2$	&	$<2$	&	$-0.01(3)$	\\
      &$\overline{K}^*_1 N$	&	$<2$	&	$<1$	&	$-0.02(5)$	&  &  $\overline{K}^*_1 N$	&	$15(6)$	&	$8(4)$	&	$+0.09(3)$		 \\
      &$(\overline{K}^*_3 N)_D$	&	$<12$	&	$<3$	&	$-0.09(5)$	& & $(\overline{K}^*_3 N)_D$	&	$<9$	&	$<5$	&	$-0.04(4)$\\
  & &  & & &                                                                                                             & &&&\\
    $\Sigma(1900) S_{11}$  &  $\overline{K} N$ 	&	$127(51)$	&	$67(17)$&	$+0.67(17)$	   &    $\Sigma(2000) S_{11}$  & $\overline{K} N$ 	&	$<10$	&	$<4$	&	$+0.01(3)$\\
        $1900(21)$ & $\pi \Lambda$ 	&	$<11$	&	$<6$	&	$-0.13(7)$	  &                                                    $2060(20)$ & $\pi \Lambda$ 	&	$<2$	&	$<1$	&	$+0.00(1)$\\
         $191(47)$  & $\pi \Sigma$ 	&	$18(11)$	&	$10(5)$	&	$+0.25(8)$                                         &    $300(134)$ &$\pi \Sigma$ 	&	$28(12)$		&	$9(3)$	&	$-0.03(4)$\\
$(New)$& $(\pi \Sigma^*(1385))_D$	&	$<12$	&	$<6$	&	$+0.13(9)$	&	$\ast$&$(\pi \Sigma^*(1385))_D$	&	$<2$	&	$<1$	&	$+0.00(1)$ \\
&$(\pi \Lambda^*(1520))_P$	&	$<15$	&	$<8$	&	$-0.14(11)$	 &&	$(\pi \Lambda^*(1520))_P$	&	$<4$	&	$<2$	&	$-0.00(2)$	\\
&$(\overline{K} \Delta)_D$	&	$<1$	&	$<1$	&	$+0.00(11)$	&&	$(\overline{K} \Delta)_D$	&	$<18$	&	$<7$	&	$+0.01(3)$	\\
&$\overline{K}^*_1 N$	&	$<15$&	$<8$	&	$+0.13(12)$	&&$\overline{K}^*_1 N$	&	$<9$	&	$<3$	&	$-0.01(2)$\\
&$(\overline{K}^*_3 N)_D$	&	$<24$	&	$<13$	&	$+0.15(18)$&	&$(\overline{K}^*_3 N)_D$	&	$<5$	&	$<2$	&	$-0.00(4)$	\\
   & &  & & &                                                                                                             & &&&\\
      $\Sigma(1770) P_{11}$  &  $\overline{K} N$ 	&	$<1$	&	$<1$	&	$+0.00(1)$	         &    $\Sigma(1880) P_{11}$  &$\overline{K} N$ 	&	$30(8)$ &	$10(3)$	&	$+0.10(3)$	 \\
      $1732(28)$  & $\pi \Lambda$ 		&	$<1$	&	$<1$	&	$-0.00(1)$			         &    $1821(17)$  &  $\pi \Lambda$ 		&	$<5$	&	$<2$	&	$-0.02(2)$	               \\
       $200(41)$   & $\pi \Sigma$ 		&	$64(21)$		&	$32(10)$	&	$-0.01(6)$	     &             $300(59)$&  $\pi \Sigma$ 		&	$<3$	&	$<1$	&	$-0.01(3)$	  \\   
   $\ast$& $(\pi \Sigma^*(1385))_P$	&	$84(33)$	&	$42(13)$	&	 $-0.01(6)$	             &                  $\ast\ast$          &  $(\pi \Sigma^*(1385))_P$	&	$<10$	&	$<4$	&	$-0.03(3)$ \\   
 &$(\pi \Lambda^*(1520))_D$	&	$<11$		&	$<6$	&	$-0.00(1)$&		 & $(\pi \Lambda^*(1520))_D$	&	$5(5)$	&	$2(1)$	&	$-0.04(2)$		\\
&$(\overline{K} \Delta)_P$	&	$23(19)$		&	$11(9)$	&	 $+0.00(3)$&	   & $(\overline{K} \Delta)_P$	&	$116(35)$		&	$39(8)$	&	$+ 0.20(2)$		\\
&$\overline{K}^*_1 N$ 	&	$14(8)$		&	$7(4)$	&	$+0.00(3)$&	 &  $\overline{K}^*_1 N$ 	&	$<1$	&   $<1$	&	$+0.00(2)$	\\
&$(\overline{K}^*_3 N)_P$	&	$10(5)$	&	$5(2)$		&	$+0.00(2)$&	&  $(\overline{K}^*_3 N)_P$	&	$<10$	&	$<3$	&	$+0.03(2)$		 \\
    
        \end{tabular}
      \end{ruledtabular}
      \end{center}
      \end{table*}

\begin{table*}[htpb]
\addtocounter{table}{-1}
 \caption {Cont'd.}  
\begin{center}
\begin{ruledtabular}
 \begin{tabular}{cccccccccc}    
   Resonance     &Channel & $\Gamma_i$ (MeV) & ${\cal B}_i$ ($\%$) & $\sqrt{xx_i}$ &{Resonance}    & {Channel} & {$\Gamma_i$ (MeV)} & {${\cal B}_i$ ($\%$)}  & {$\sqrt{xx_i}$ }\\ \hline

        $\Sigma(1730) P_{13}$   & $\overline{K} N$ 	&	$5(4)$		&	$2(1)$	&	$+0.02(1)$         &       $\Sigma(1940) P_{13}$ &  $\overline{K} N$ 	&	$53(10)$	&	$13(2)$	&	$+0.13(2)$ \\
        $1727(27)$ & $\pi \Lambda$	 	&	$194(65)$	&	$70(17)$	&	$+0.11(3)$     &        $1941(18)$     & $\pi \Lambda$		&	$<7$	&	$<2$	&	$+0.03(3)$	 \\
        $276(87)$   &$\pi \Sigma$ 		&	$32(19)$	&	$12(6)$	&	$+0.04(2)$     &      $400(49)$&  $\pi \Sigma$ 		&	$17(10)$	&	$4(2)$	&	$+0.07(2)$	\\     
 $(New)$& $(\pi \Sigma^*(1385))_P$	&	$<42$		&	$<6$	&	+0.03(4)   &     $(New)$  & $(\pi \Sigma^*(1385))_P$	&	$89(33)$	&	$22(7)$	&	$+0.17(3)$	 	\\
& $(\pi \Sigma^*(1385))_F$	&	$<2$	&	$<1$	&	$-0.00(3)$		&       & $(\pi \Sigma^*(1385))_F$	&	$<2$	&	$<1$	&	$+0.01(4)$ \\
&$(\pi \Lambda^*(1520))_S$	&	$<5$	&	$<2$	&	$+0.01(3)$	         & &$(\pi \Lambda^*(1520))_S$	&	$18(10)$		&	$5(2)$		&	$+0.08(2)$		\\
&$(\pi \Lambda^*(1520))_D$	&	$<9$	&	$<3$	&	$-0.01(2)$		&&$(\pi \Lambda^*(1520))_D$	&	$<10$	&	$<3$	&	$+0.03(4)$		\\
&$(\overline{K} \Delta)_P$	&	$<28$&	$<11$	&	$-0.03(2)$	 	&&$(\overline{K} \Delta)_P$	&	$<10$	&	$<3$	&	$+0.03(3)$		\\
&$\overline{K}^*_1 N$ 	&	$<7$	&	$<3$	&	$+0.01(1)$	&	 &$\overline{K}^*_1 N$ 	&	$<2$	&	$<1$	&	$+0.00(4)$	 \\
&$(\overline{K}^*_3 N)_P$	&	$<28	$&	$<10$	&	$-0.03(2)$	 &  &	$(\overline{K}^*_3 N)_P$	&	$<20$	&	$<5$	&	$+0.05(4)$	 	\\
   & &  & & &                                                                                                             & &&&\\
        $\Sigma(1670) D_{13}$  &  $\overline{K} N$ 	&	$3.4(4)$	&	$6.2(7)$	&	$+0.06(1)$	&$\Sigma(1775) D_{15}$    &$\overline{K} N$	&	$52(1)$	&	$40(1)$	&	$+0.40(1)$\\
         $1678(2)$ & $\pi \Lambda$ 		&	$6(1)$		&	$11(2)$	&	$+0.08(1)$                               & $1778(1)$ & $\pi \Lambda$		&	$33(1)$	&	$25(1)$	&	$-0.31(1)$ \\
         $55(4)$ & $\pi \Sigma$ 		&	$34(4)$	&	$62(7)$	&	$+0.20(1)$	                                    &$131(3)$ & $\pi \Sigma$		&	$2.0(3)$		&	$1.6(3)$	&	$+0.08(1)$	 \\
        $\ast\ast$$\ast\ast$& $(\pi \Sigma^*(1385))_S$ 	&	$<3$	&	$<5$	&	$-0.01(9)$ &   $\ast\ast$$\ast\ast$  & $(\pi \Sigma^*(1385))_D$	&	$5(1)$		&	$4(1)$&	$-0.12(1)$            \\      
          &$(\pi \Sigma^*(1385))_D$ 	&	$<11$	&	$<20$	&	$+0.06(6)$	 	&& $(\pi \Sigma^*(1385))_G$	&	$<1$	&	$<1$	&	$+0.02(1)$	\\
&$(\pi \Lambda^*(1520))_P$	&	$<1$	&	$<1$	&	$+0.01(3)$		&& $(\pi \Lambda^*(1520))_P$	&	$1.0(5)$	&	$0.8(4)$	&	$-0.06(1)$		\\
&$(\pi \Lambda^*(1520))_F$	&	$<1$	&	$<1$	&	$-0.01(1)$		&& $(\pi \Lambda^*(1520))_F$	&	$<1$	&	$<1$&	$+0.01(1)$	\\
&$(\overline{K} \Delta)_S$	&	$<12$	&	$<22$	&	$+0.05(11)$	 	&& $(\overline{K} \Delta)_D$	&	$1(1)$	&	$1(1)$	&	$+0.06(3)$	\\
&$(\overline{K} \Delta)_D$	&	$<5$	&	$<8$	&	$+0.02(9)$   	&& $\overline{K}^*_1 N$	&	$0.5(2)$	&	$0.4(2)$	&	$+0.04(1)$	\\
&$\overline{K}^*_1 N$ 	&	$<2$	&	$<3$	&	$-0.02(3)$	                  && $(\overline{K}^*_3 N)_D$	&	$0.5(2)$	&	$0.4(2)$	&	$+0.04(1)$	 \\
&$(\overline{K}^*_3 N)_S$	&	$<5$	&	$<9$	&	$+0.04(4)$	  && &&&\\
&$(\overline{K}^*_3 N)_D$	&	$<2$&	$<3$	&	$+0.02(3)$	 &&&&&	\\
   & &  & & &                                                                                                             & &&&\\
       $\Sigma(1915) F_{15}$  &  $\overline{K} N$	&	$3.9(9)$	&	$2.6(4)$	&	$+0.026(4)$&    $\Sigma(2030) F_{17}$  &  $\overline{K} N$		&	$27(3)$	&	$13(1)$	&	$+0.13(1)$	\\
        $1920(7)$& $\pi \Lambda$		&	$<3$	&	$<2$	&	$-0.01(1)$	  &    $2030(5)$& $\pi \Lambda$			&	$36(4)$	&	$17(2)$	&	$+0.15(1)$	 \\
               $149(17)$& $\pi \Sigma$		&	$113(18)$	&	$76(12)$	&	$-0.14(1)$          &    $207(17)$& $\pi \Sigma$			&	$11(2)$	&	$5(1)$	&	$-0.08(1)$		\\
   $\ast\ast$$\ast\ast$    & $(\pi \Sigma^*(1385))_P$	&	$<1$	&	$<1$	&	$-0.00(2)$       & $\ast\ast$$\ast\ast$ & $(\pi \Sigma^*(1385))_F$	&	$38(7)$ 	&	$18(3)$	&	$+0.16(1)$                \\
& $(\pi \Sigma^*(1385))_F$	&	$20(14)$	&	$13(9)$	&	$+0.06(2)$		&& $(\pi \Lambda^*(1520))_D$	&	$<2$	&	$<1$	&	$-0.02(1)$	\\
&$(\pi \Lambda^*(1520))_D$	&	$<2$	&	$<1$	&	$-0.01(2)$	&&$(\pi \Lambda^*(1520))_G$	&	$<1$	&	$<1$	&	$-0.00(1)$	\\
&$(\pi \Lambda^*(1520))_G$	&	$<2$	&	$<1$	&	$-0.01(1) $    &&$(\overline{K} \Delta)_F$	&	$24(9)$	&	$12(4)$	&	$+0.12(2)$	\\
&$(\overline{K} \Delta)_P$	&	$<2$	&	$<1$	&	$-0.00(2)$	&&$\overline{K}^*_1 N$		&	$6(3)$	&	$3(1)$		&	$+0.06(2)$	\\
&$\overline{K}^*_1 N$		&	$<1$	&	$<1$	&	$-0.00(1)$	&&$(\overline{K}^*_3 N)_F$	&	$3(2)$	&	$2(1)$		&	$+0.05(1)$	\\
&$(\overline{K}^*_3 N)_P$	&	$<1$	&	$<1$	&	$+0.00(2)$	&&$(\overline{K}^*_3 N)_H$	&	$<2$	&	$<1$	&	$+0.02(1)$	\\
&$(\overline{K}^*_3 N)_F$	&	$7(5)$	&	$5(3)$	&	$+0.04(1)$	&&  &&&	\\
      \end{tabular}
      \end{ruledtabular}
      \end{center}
      \end{table*}
 The pole positions in the $S$- or $T$-matrix for each partial wave differ from Breit-Wigner parameters because the total widths are energy dependent. For example, if a $T$-matrix amplitude varied as $T\sim (M-W-{\rm i}\Gamma(W)/2)^{-1}$, then the Breit-Wigner mass and width would be $M$ and $\Gamma(M)$, respectively, whereas the complex pole position $W_p$ would be the energy where 
 $M-W_p-{\rm i}\Gamma(W_p)/2 = 0 $.
 Tables III and IV list the complex $S$-matrix pole positions of the resonances for partial waves $I=0$ and $I=1$, respectively. The first column lists the resonance name. the second column shows the rating of the resonance (in terms of the number of stars) \cite{pdg12}. The third column lists the pole mass, which is the real part of the pole position. The pole width, given by twice the negative of the imaginary part, is listed in the fourth column. 
To the best of our knowledge, pole positions for these $\Lambda^*$ and $\Sigma^*$ states have not been extracted from any prior multichannel analyses.
   \begin{table*}[htpb]
 \caption { $S$-matrix pole positions (in MeV) for $I=0$ states from this work.}
 \begin{center}
 \begin{ruledtabular}
  \begin{tabular}{ccccccccc}
  Resonance             &   Rating             &Real Part  &     $-2\times$Imaginary Part    &     Resonance             &  Rating       & Real Part  &     $-2\times$Imaginary Part    \\
\hline
           $\Lambda$(1670)$S_{01}$   & $ \ast\ast$$\ast\ast$     &  $1667$    &    $26$        &                             $\Lambda(1520)D_{03}$    &$\ast\ast$$\ast\ast$         &$1518$       &$16$                                   \\
         &&&&&&&\\
           $\Lambda$(1800)$S_{01}$   &$\ast$$\ast\ast$       & $1729$   &  $198$                        &     $\Lambda(1690)D_{03}$    &$\ast\ast$$\ast\ast$         &$1689$       &$53$ \\
           &&&&&&&\\
           $\Lambda(2000)S_{01}$       &$New$            &$1984$       &$233$          &          $\Lambda(2050)D_{03}$    &$New$       &$1985$       &$447$                                                              \\
        &&&&&&&\\
            $\Lambda(1600)P_{01}$  &$\ast$$\ast\ast$        & $1572$          & $138$                          &     $\Lambda(1830)D_{05}$     &$\ast\ast$$\ast\ast$         &$1809$      &$109$     \\
           &&&&&&&\\
            $\Lambda(1710)P_{01}$    & $New$  &$1688$          &$166$                  &             $\Lambda(1820)F_{05}$     &$\ast\ast$$\ast\ast$        &$1814$      &$85$     \\
          &&&&&&&\\
             $\Lambda(1810)P_{01}$  &$\ast$$\ast\ast$           &$1780$       &$64$     &       $\Lambda(2110)F_{05}$    &$\ast$$\ast\ast$           &$1970$     &$350$   \\
       &&&&&&&\\
              $\Lambda(1890)P_{03}$      &$\ast\ast$$\ast\ast$         &$1876$      &$145$ & $\Lambda(2020)F_{07}$     &$\ast$            &$1999$      &$146$             &                    \\
  &&&&&&& \\  
              &&&&$\Lambda(2100)G_{07}$    &$\ast\ast$$\ast\ast$        &$2023$     &$239$                     \\
            \end{tabular}
            \end{ruledtabular}
            \end{center}
            \end{table*}

    \begin{table*}[htpb]
 \caption { $S$-matrix pole positions (in MeV) for  $I=1$ states from this work.}
 \begin{center}
 \begin{ruledtabular}
  \begin{tabular}{ccccccccc}
  Resonance             &   Rating             &Real Part  &     $-2\times$Imaginary Part    &     Resonance             &  Rating       & Real Part  &     $-2\times$Imaginary Part    \\
\hline
           $\Sigma$(1620)$S_{11}$   &  $\ast\ast$    &  $1501$    &    $171$        &                             $\Sigma(1730)P_{13}$  &$New$         &$1683$            & $243$                                 \\
         &&&&&&&\\
           $\Sigma$(1750)$S_{11}$   &$\ast$$\ast\ast$       & $1708$   &  $158$                        &     $\Sigma(1940)P_{13}$      &$New$         &$1874$      &$349$   \\
           &&&&&&&\\
           $\Sigma(1900)S_{11}$       &$New$            &$1887$       &$187$          &         $\Sigma(1670)D_{13}$    &$\ast\ast$$\ast\ast$         &$1674$       &$54$                                             \\
        &&&&&&&\\
            $\Sigma(2000)S_{11}$  &$\ast$        & $2040$          & $295$                          &     $\Sigma(1775)D_{15}$    &$\ast\ast$$\ast\ast$         &$1759$       &$118$      \\
           &&&&&&&\\
            $\Sigma(1770)P_{11}$    & $\ast$  &$1693$          &$163$                  &             $\Sigma(1915)F_{15}$    &$\ast\ast$$\ast\ast$       &$1897$       &$133$      \\
          &&&&&&&\\
             $\Sigma(1880)P_{11}$  &$\ast\ast$           &$1776$       &$270$     &       $\Sigma(2030)F_{17}$     &$\ast\ast$$\ast\ast$        &$1993$      &$176$   \\
            \end{tabular}
            \end{ruledtabular}
            \end{center}
            \end{table*}


\section{COMPARISONS WITH QUARK-MODEL PREDICTIONS}

 In Tables V and VI we present decay amplitudes for various channels and compare our results for $I=0$ and $I=1$ states, respectively, with a quark model. The magnitude of the decay amplitude is equal to $\sqrt{\Gamma_i}$~, the square root of the partial width for the channel. Its sign is the phase relative to the $\overline K N$ coupling (taken to be positive). The values in the first row are our results, while those in the second row are from Koniuk and Isgur \cite{koniuk80}. The channels included are $\overline K N$, $\pi\Sigma$, $\pi\Lambda$, $\pi\Sigma^*(1385)$, and $\overline K\Delta$. The subscript `l' or `h' that appears with a channel represents the lower or higher orbital angular momentum of that channel. 

 
 
For $I=0$ states, our $\overline K N$ results are all in good agreement with model values except for $\Lambda(2050)D_{03}~(New)$. For this state, our elastic coupling of 9.8 is larger than the model  value of 1.1.

 Our $\pi \Sigma$ results are in good agreement with the model predictions for  all states except $\Lambda(1710)P_{01}$ $(New)$, $\Lambda(1800)S_{01}$, and $\Lambda(2050)D_{03}~(New)$.
 Our values for these states disagree either in sign or magnitude with the predictions. For $\Lambda(1800)S_{01}$ we agree with the predicted signs of all couplings.
 
 Our $\eta \Lambda$ results agree very well with the quark model after the model predictions were scaled by an overall sign of $-1$. For $\Lambda(1670)S_{01}$ and $\Lambda(1800)S_{01}$ our $\eta\Lambda$ values were $-3.1$ and +3.8 compared with the model values of $-2.2$ and $-3.9$, respectively. 
 
 For $I=1$ states, our $\overline KN$ results are in good agreement with model values except for $\Sigma(1620)S_{11}$, $\Sigma(1880)P_{11}$, $\Sigma(1900)S_{11}~(New)$, and $\Sigma(1940)P_{13}~(New)$. For these states our elastic amplitudes of 15, 5.5, 11, and 7.3 are larger than the corresponding model values of 5.3, 1.8, 2.5, and 0.3, respectively.
 
Our $\pi \Sigma$ results are in good agreement with the model predictions for  all states except $\Sigma(1750)S_{11}$, $\Sigma(1730)P_{13}~(New)$,  and $\Sigma(1880)P_{11}$.  
 For these states our $\pi\Sigma$ amplitudes disagree with model values both in magnitude and sign. For $\Sigma(1900)S_{11}$ our value disagrees in sign with the model value while for the state $\Sigma(1940)P_{13}~(New)$ our value of +4 disagrees in magnitude with model value of +0.7.

\begin{table*}[htpb]
\caption{Comparison of decay amplitudes for $I=0$ states with predictions of quark models.  The first row gives our results, while the second row lists the available $\overline K N$, $\pi\Sigma$, $(\pi\Sigma^*(1385))_l$, and $(\pi\Sigma^*(1385))_h$ amplitudes predicted by Koniuk and Isgur \cite{koniuk80}.} 
\begin{center}
\begin{ruledtabular}
\begin{tabular}{ccccc}
State   & $\overline K N$   & $\pi\Sigma$  &   $(\pi\Sigma^*(1385))_l$    & $(\pi\Sigma^*(1385))_h$  \\ \hline
$\Lambda(1405)S_{01}$   &-- & +7.1  &  --          & --   \\
    $\ast \ast$$ \ast \ast$                           &--   &$+7.4$ & --   & --   \\
 
 $\Lambda(1670)S_{01}$  &$2.7(7)$     &$-3.0(5)$   &--&$-0.6(5)$\\
     $\ast \ast $$\ast \ast$                          &$3.3$           &$-3.2$   &--       &$-1.2$ \\
                              
$\Lambda(1800)S_{01}$  & $5.8(6)$   &$-3.1(6)$ &--& $-3.9(9)$  \\
        $\ast \ast$$\ast$                       & $2.9$         &$-11$     &--   &$-5.5$\\ 

$\Lambda(1600) P_{01}$  &$4.6(5)$     &$-7.5(7)$  &--&$-1.3(17)$ \\
   $\ast \ast $$\ast$                            &$5.4$         &$-3.8$ &--&$-2.1$ \\

$\Lambda(1710)P_{01}$  &$8.8(7)$     &$-6.1(8)$  &--&$+6.0(8)$ \\
     $New$                          &$5.7$            &$+6.0$           &--    &$+1.6$  \\

$\Lambda(1810)P_{01}$  &$5.8(6)$     &$-2.4(6)$  &--&$+0.3(7)$\\ 
       $\ast$                       &$4.6$          &$-3.8$            &-- &$+4.2$  \\

$\Lambda(1890)P_{03}$  & $7.7(5)$   &$-1.8(5)$ & $-1.2(9)$ & $-6.5(7)$ \\
        $\ast \ast$$\ast\ast$                       & $7.4$         &$-2.1$       &$-0.1$ & $ -1.1$\\

$\Lambda(1520)D_{03}$  & $2.78(4)$   &$+2.8(1)$ &$-0.7(7)$&$ -0.2(4)$ \\
   $\ast \ast$$ \ast \ast$                            &$3.0$   &$+2.8$ &+small & +small     \\



$\Lambda(1690)D_{03}$  &$3.7(2)$     &$-4.0(3)$  &$-4.1(5)$ &$-1(2)$\\
     $\ast \ast$$ \ast \ast$                          &$4.3$      &$-6.6$                &$+5.5$ & $+2.3$\\



$\Lambda(2050)D_{03}$  & $9.8(7)$   &$+5.5(9)$ & $-6(1)$&$-4(1)$ \\
     $New$                          & 1.1 &  $-5.3$ &$+14$&$ -7.7$  \\

$\Lambda(1830)D_{05}$   &$2.2(1)$  & $-6.9(5)$   & $+7.7(5)$ & $ -1.5(8)$     \\
       $\ast \ast$$ \ast\ast$                         &$1.5$       &$-7.7$      & $-7.8$ & $ 0.0$         \\

 $\Lambda(1820)F_{05}$   &$6.91(8)$  & $-3.6(1)$   &  $-2.6(2)$ & $+0.7(2)$    \\
     $\ast \ast$$ \ast \ast$                           &$6.4$       & $-2.0$           &$+1.5$ & $-0.5$  \\
                      

$\Lambda(2110)F_{05}$  &$5.8(3)$     &$+2.8(5)$ &$+2.7(6)$ &  $-1.2(7)$\\
        $\ast \ast$$\ast$                       &1.8         &$+7.4$     & $+0.4$ & $ -0.4$    \\
        
        $\Lambda(2020)F_{07}$  &$2.4(5)$    &$+2.0(6)$&--   &--\\
        $\ast$                      &$1.7$         &$+4.0$    &$+4.1$  &--\\

 \end{tabular}
\end{ruledtabular}
\end{center}
\end{table*}

\begin{table*}[htpb]
\caption{Comparison of decay amplitudes for $I=1$ states with predictions of quark models.  The first row gives our results, while the second row lists the available $\overline K N$, $\pi\Sigma$, $\pi\Lambda$, ($\pi\Sigma^*(1385))_l$, ($\pi\Sigma^*(1385))_h$, $(\overline K\Delta)_l$ and $(\overline K\Delta)_h$ amplitudes predicted by Koniuk and Isgur \cite{koniuk80}. Both $\Sigma(1900)S_{11}$ and $\Sigma(2000)S_{11}$ are compared with Koniuk and Isgur's model state at 1810 MeV.}  
\begin{center}
\begin{ruledtabular}
\begin{tabular}{cccccccc}
State   & $\overline K N$   & $\pi\Sigma$  &   $\pi\Lambda$    & ($\pi\Sigma^*(1385))_l$ & ($\pi\Sigma^*(1385))_h$    & $(\overline K\Delta)_l$ & $(\overline K\Delta)_h$\\ \hline
                               
$\Sigma(1620)S_{11}$  &$15(2)$     &$+8.3(9)$  &$+4(1)$  &--&$+4(1)$&--&$+1.8(7)$\\
   $\ast \ast $                           &$5.3$         &$+9.9$  &$0.0$&--&$-0.1$  &--&--\\
$\Sigma(1750)S_{11}$  &$4(1)$     &$+7.4(7)$  &$+4.7(6)$  &--&$+8(1)$&--&$-0.5(9)$\\
     $\ast \ast$$ \ast$                          &$4.1$      &$-0.5$    &$-5.3$         &--     &$+0.4$&--&--\\

  $\Sigma(1900)S_{11}$  & $11(1)$   &$+4.3(8)$&$-2.3(7)$&--& $+2.2(9)$ &--&$+0(1)$\\
        $New$                       & $2.5$         &$-4.1$  &$+0.5$    &--   &$+7.4$&--&--\\

$\Sigma(2000)S_{11}$  & $1(1)$   &$-5.3(7)$&$+0.5(6)$&--& $+0(1)$ &--&$+2(2)$\\
       $\ast$                        & $2.5$         &$-4.1$  &$+0.5$     &--  &$+7.4$&--&--\\

$\Sigma(1770)P_{11}$  & $0(1)$   &$-8.0(9)$ &$-0.1(9)$&--& $-9(1)$ &--&$+5(1)$\\
   $\ast$                           &$1.2$   &$-3.7$  &$-2.9$&--&$+1.5$  &--    &--\\ 

$\Sigma(1880)P_{11}$   &$5.5(6)$  & $-1(1)$   &  $-1.3(7)$ &--& $-2(2)$ &--&$+11(1)$  \\
       $\ast \ast$                         &$1.8$       &$+7.1$          &  $+1.9$   &-- & $+2.5$   &--&--     \\

$\Sigma(1730)P_{13}$   & $2.2(9)$   &+6(2) &+14(2)& $+4(4)$ &$-0(3)$&$-4(2)$&--\\
        $New$                       & $3.9$         &$-2.1$  &$+3.3$       &$-8.2$&--&--&--\\

$\Sigma(1940)P_{13}$  &$7.3(7)$     &$+4(1)$  &$+1(2)$  &$+9(2)$&$+0(2)$&$+2(2)$&--\\
       $New$                        &$4.3$         &$+0.7$           &$-0.1$     & $+3.6$ &-- &--&--  \\

$\Sigma(1670)D_{13}$  &$1.9(1)$     &$+5.8(3)$  &$+2.5(2)$  &$-0(3)$&$+2(2)$&$+1(3)$&$+1(3)$\\
     $\ast \ast $$\ast \ast$                          &$2.1$          &$+6.6$       &$+2.4$       &$+0.9$ & $+0.5$&--&-- \\


$\Sigma(1775)D_{15}$  &$7.21(6)$     &$+1.4(1)$  &$-5.71(9)$  &$-2.2(3)$&$+0.3(2)$&$+1.0(5)$&--\\
     $\ast \ast$$ \ast \ast$                          &$6.7$            &$+3.0$          &$-4.7$        &$+2.9$ & $0.0$&--&--  \\


$\Sigma(1915)F_{15}$   &$2.0(2)$  & $-10.6(9)$   &  $-1.1(7)$ & $-0(1)$   &$+4(2)$&$-0(2)$&--  \\
     $\ast \ast$$ \ast \ast$                           &$1.1$       & $-5.3$      &  $-3.3$      &$-0.8$&$+0.5$ &--&-- \\
        

 $\Lambda(2030)F_{17}$  &$5.2(3)$     &$-3.4(3)$  &$+6.0(3)$  &$+6.2(6)$&--&$+4.9(9)$&--\\
   $\ast \ast$$\ast\ast$                       &$5.4$          &$-2.2$        &$+3.2$      &$-2.1$&$ 0.0$&$-3.8$&$ 0.0$  \\
                       

 \end{tabular}
\end{ruledtabular}
\end{center}
\end{table*}

  \section{Summary and Conclusions}
  This work was undertaken to determine the parameters of $\Lambda^*$ and $\Sigma^*$ resonances with masses up to about 2.1 GeV using a global multichannel fit. 
 For the first time, we explicitly include amplitudes for $\overline K N\rightarrow \overline K N$,  $\overline K N\rightarrow \pi\Lambda$, and $\overline K N\rightarrow \pi \Sigma $ in addition to those for $\overline K N\rightarrow \pi \Sigma^*$, $\overline K N\rightarrow \pi\Lambda^*$, $\overline K N\rightarrow \overline K^*N$, and $\overline K N\rightarrow \overline K \Delta$. This makes the present work the most comprehensive multichannel fit to date for $\overline KN$ scattering reactions. We found no evidence for the $\Sigma(1660)P_{11}$ or $\Sigma(1940)D_{13}$, which are rated as 3-star resonances by the Particle Data Group \cite{pdg12}. Our results on resonance parameters for most states agree very well with quark-model predictions. Furthermore, our analysis finds evidence for five new states. The proposed states are $\Lambda(1710)P_{01}$, $\Lambda(2050)D_{03}$, $\Sigma(1900)S_{11}$, $\Sigma(1730)P_{13}$, and $\Sigma(1940)P_{13}$. 
  
  
\acknowledgements{This work was supported by the U.S. Department of Energy Grant No. DE-FG02-01ER41194. 

\end{document}